\documentstyle[a4,
twoside]{article}
\makeatletter
\renewcommand{\@oddhead}{The Ground, a String, Two Elastic Springs
 \hfill \thepage}
\renewcommand{\@evenhead}{\thepage \hfill S. A. Choro\v{s}avin }
\renewcommand{\@oddfoot}{}
\renewcommand{\@evenfoot}{}
\makeatother
\newenvironment{Thm}[2]
{\par\addvspace{\bigskipamount}\noindent {\bf #1#2}\it }%
{\par\addvspace{\bigskipamount}\noindent }

\newenvironment{Remark}[1]{\begin{Thm}{Remark}{#1}}{\end{Thm} }

\author{S.A.~Choro\v{s}avin}
\title{ 
 The Ground, a String, Two Elastic Springs.
 \bigskip\\
  Simple Exactly Solvable Models
  of One-Dimensional Scalar Fields 
  with Concentrated Factors. 
  }
\date{}
\begin{document}
\maketitle 
\begin{abstract}
 This paper is a basis of a part of my set of lectures, 
 subject: `Formal methods in solving differential equations and 
 constructing models of physical phenomena'. Addressed, mainly: 
 postgraduates and related readers. 
 Content: 
 a discussion of the simple models
 ( I would rather say, toy models )
 of the interaction based 
 on equation arrays of the kind: 
\begin{eqnarray*}
 \frac{\partial^2 u(t,x)}{\partial t^2}
 &=&
 c^2\frac{\partial^2 u(t,x)}{\partial x^2}
                 -4{\gamma_a}c\delta(x-x_a) Q_a(t)
                 -4{\gamma_b}c\delta(x-x_b) Q_b(t), 
\\
 Q_a(t)
 &=& u(t,x_a)
\\
 Q_b(t)
 &=& u(t,x_b)
\\
\end{eqnarray*}
 Central mathematical points: d'Alembert-Kirchhoff-like Formulae,
 Finite Rank Perturbations.
\end{abstract}

\newpage
\section*%
{ Introduction }

 As long as the researcher takes an interest in resolvent formulae 
 of finite rank perturbed operators, in other words, 
 as long as he or she, the researcher,
 prefers to remain in the space-FREQUENCE framework,
 so long he has an infinite series of good papers, articles, books,
 manuals...

 But as soon as
 the researcher is turning his attention to the problems of the "space-TIME", 
 in other words, as soon as
 he has need to solve the associated wave equation,
 as soon as he has need of a suitable d'Alembert-like formula, or so, - 
 in that moment the situation is changing dramatically. 
  
 I cannot say "there is no paper on the subject at all",
 I cannot say "I have not seen it",
 but if anyone asks me "where have you seen it?, where is this `there'?",
 I will become very "pensativo",
 and I am in doubt that I will be "solitario" in this state.

 So, I tried, try and (I hope) will try to collect 
 the suitable examples of simple exactly solvable models of wave'n'particle.

 Whether my collection is worth to discuss it, you solve.
 

\section%
{ Models of Two-Point Interaction with 
  an only one-dimensional Scalar Field }

\subsection%
{ Preliminaries }
 In this paper we fix measure units 
 and let
$x$ 
 be dimensionless position parameter, i.e., 
$$
 \mbox{ physical position coordinate }
  =  [\mbox{ length unit }]\times x +const \,.
$$
 Otherwise a confusion can ocurr, in relating to the definition 
$$
 \int_{-\infty}^{\infty}\delta(x-x_0)f(x)dx=f(x_0) \,.
$$
 We assume the standard foramalism, where 
$$ 
 \delta(x-x_0) = \frac{\partial 1_{+}(x-x_0)}{\partial x}
$$
 and where 
$1_{+}$
 stands for a unit step function (Heaviside function):
$$ 
 1_{+}(\xi) := 1(\xi \geq 0) := \Bigl\{
 \begin{array}{ccc}
 1&,&\mbox{ if } \xi \geq 0 \,,
 \\
 0&,&\mbox{ if } \xi < 0 \,.
 \\
 \end{array}
$$

\bigskip

 Another feature of the notations is this. 
 We will handle the functions which have a special variable,
$t$,
 that means no doubt "time", and we will be interested in the case where 
$t \geq 0 $.
 So, we could consider the restrictions of these functions, 
 onto positive
 $t$-half-line. 
 But it will be {\bf technically } more convenient to
 {\bf redefine } the functions,
 putting them zero on negative
 $t$-half-line.
 For more details of this feature of the notations see 
 {\bf The 
$ \cdot 1_{+}()$
 Convention },
 in the next subsection.

\bigskip

 A few words about the models:
 Recently I have already presented some models of 
 ONE particle of FINITE mass, 
 interacting with scalar field,
 and the interaction has been concentrated at one, only ONE, point. 
 
 Now I discuss models of TWO-point concentrated interaction, but the mass 
 of the particle (or particles) I assume to be INFINITE,
 or, more precisely, I assume the particle(s)
 to be of infinite mass and immobile: motionless, fixed. 

 A naive formulation of the situation is: 
 let us look at the Ground (the infinite immobile mass),
 the Earth or the Moon say,
 connected with 
 a String (very-very long thin tensed cord) by means of two Ideal
 (we never approve Imperfection, don't we?) Springs.
 What will we tell then, ideally?

\newpage
\subsection%
{ D'Alembert-Kirchhoff-like formulae }

 Recall that a standard D'Alembert-Kirchhoff-like formula reads: 
 if 
$$
 \frac{\partial^2 u}{\partial t^2}
  = c^2\frac{\partial^2 u}{\partial x^2} + f
 \,,\quad
 u=u(t,x)
 \,,\quad
 f=f(t,x)
 \,,\quad
 (t\geq 0)
           \eqno(*)
$$
$$
  f=-4{\gamma_a}c\delta(x-x_a)\Big(F_{src\ a}(t)\Big)
    -4{\gamma_b}c\delta(x-x_b)\Big(F_{src\ b}(t)\Big)
 \qquad x_a \leq x_b
$$
 and given initial data, 
$u(0,\cdot )$ 
 and 
$\frac{\partial u(t,\xi)}{\partial t})\Big|_{t=0}$,
 then, for 
$t\geq 0$, 

\medskip

\begin{eqnarray*}
 u(t,x)
 &=&
 \displaystyle
 -{2\gamma_a}
 \int_0^{t-|x-x_a|/c}
 \Big(F_{src\ a}(\tau)\Big)d\tau \cdot 1_{+}(t-|x-x_a|/c)
 \\&&{}
 -{2\gamma_b}
 \int_0^{t-|x-x_b|/c}
 \Big(F_{src\ b}(\tau)\Big)d\tau \cdot 1_{+}(t-|x-x_b|/c)
 \\[\medskipamount]\\
 &&{}\qquad + u_{0}(t,x)
 \\[\smallskipamount] \\
\end{eqnarray*}
 where 
\begin{eqnarray*}
 u_{0}(t,x)
 &:=& 
 c_{+}(x+ct) + c_{-}(x-ct)
 \\&=&
 \frac12 \Big(u(0,x+ct) + u(0,x-ct)\Big)
 \\&&{}
   + \frac{1}{2c}\Big(
  \widetilde{\dot u}(0,x+ct) - \widetilde{\dot u}(0,x-ct)\Big)
 \\&&{}
\end{eqnarray*}
 and where 
$\widetilde{\dot u}$ 
 stands for any function defined by 
$$
 \frac{\partial\widetilde{\dot u}(0,\xi)}{\partial \xi}
   =\Bigl(\frac{\partial u(t,\xi)}{\partial t}\Bigr)\Big|_{t=0}
\,.
$$
\footnote{
 Note that
$$
\widetilde{\dot u}(0,x+ct) - \widetilde{\dot u}(0,x-ct)
$$
 does not depend on what the primitive is which one has chosen!!! 
 Moreover, we need only 
$ \widetilde{\dot u}|_{t=0} $
 and not 
$ \dot u|_{t=0}$
 itself!
} 

 If we put 
\begin{eqnarray*}
 F_a(t)
 &:=&
 \displaystyle
 -{2\gamma_a}
 \int_0^{t}
 \Big(F_{src\ a}(\tau)\Big)d\tau \cdot 1_{+}(t) \,,
\\
 F_b(t)
 &:=&
 -{2\gamma_b}
 \int_0^{t}
 \Big(F_{src\ b}(\tau)\Big)d\tau \cdot 1_{+}(t) \,,
\end{eqnarray*}
 for short, then 
\newpage\noindent

\medskip\noindent
\fbox{
\parbox{\textwidth}{
\begin{eqnarray*}
 u(t,x)
 &=&
 F_{a}(t-|x-x_a|/c)
 + F_{b}(t-|x-x_b|/c)
 + u_{0}(t,x)
\qquad (t \geq 0)
\\
\end{eqnarray*}
} 
} 

\medskip\noindent
 Another and a little more correct form of this expression is:

\medskip\noindent
\fbox{
\parbox{\textwidth}{
\begin{eqnarray*}
 u(t,x) \cdot 1_{+}(t)
 &=&
 F_{a}(t-|x-x_a|/c)
 + F_{b}(t-|x-x_b|/c)
 + u_{0}(t,x) \cdot 1_{+}(t)
\\
\end{eqnarray*}
} 
} 

\medskip\noindent
 Now we set 
$$
 T := |x_a-x_b|/c
$$
 and concentrate on 
$$
 x = x_a , x_b
$$ 
 by writing 
%
\begin{eqnarray*}
 u(t,x_a)
 &=&
 F_{a}(t)
 + F_{b}(t-T)
 + u_{0}(t,x_a)
\qquad (t \geq 0)
\\
\end{eqnarray*}
\begin{eqnarray*}
 u(t,x_b)
 &=&
 F_{a}(t-T)
 + F_{b}(t)
 + u_{0}(t,x_b)
\qquad (t \geq 0)
\\
\end{eqnarray*}
 We transform these expressions into 
%
%
\par\medskip\noindent
\fbox{
\parbox{\textwidth}{
\begin{eqnarray*}
 F_{a}(t)
 &=&
 \Bigl(
 u(t,x_a)
 - u_{0}(t,x_a)
 - F_{b}(t-T)
 \Bigr) \cdot 1_{+}(t)
\\
\end{eqnarray*}
\begin{eqnarray*}
 F_{b}(t)
 &=&
 \Bigl(
 u(t,x_b)
 - u_{0}(t,x_b)
 - F_{a}(t-T)
 \Bigr) \cdot 1_{+}(t)
\\
\end{eqnarray*}
} 
} 

\medskip\noindent
 and then 
%
\begin{eqnarray*}
 F_{a}(t)
 &=&
 \Bigl( u(t,x_a) - u_{0}(t,x_a) \Bigr) \cdot 1_{+}(t)
\\&&{}
 -
 \Bigl(
 u(t-T,x_b) - u_{0}(t-T,x_b)
 \Bigr) \cdot 1_{+}(t-T)
\\&&{}
 + F_{a}(t-2T) \cdot 1_{+}(t-T)
\\[\smallskipamount]\\
 F_{b}(t)
 &=&
 \Bigl( u(t,x_b) - u_{0}(t,x_b) \Bigr) \cdot 1_{+}(t)
\\&&{}
 -
 \Bigl(
 u(t-T,x_a) - u_{0}(t-T,x_a)
 \Bigr) \cdot 1_{+}(t-T)
\\&&{}
 + F_{b}(t-2T) \cdot 1_{+}(t-T)
\\
\end{eqnarray*}
 \newpage\noindent
\begin{eqnarray*}
 F_{a}(t)
 &=&
 \Bigl( u(t,x_a) - u_{0}(t,x_a) \Bigr) \cdot 1_{+}(t)
\\&&{}
 -
 \Bigl(
 u(t-T,x_b) - u_{0}(t-T,x_b)
 \Bigr) \cdot 1_{+}(t-T)
\\&&{}
 +
 \Bigl(
 u(t-2T,x_a) - u_{0}(t-2T,x_a)
 \Bigr) \cdot 1_{+}(t-2T)
\\&&{}
 -
 \Bigl(
 u(t-3T,x_b) - u_{0}(t-3T,x_b)
 \Bigr) \cdot 1_{+}(t-3T)
\\&&{}
 + F_{a}(t-4T) \cdot 1_{+}(t-3T)
\\[\smallskipamount]\\
 F_{b}(t)
 &=&
 \Bigl( u(t,x_b) - u_{0}(t,x_b) \Bigr) \cdot 1_{+}(t)
\\&&{}
 -
 \Bigl(
 u(t-T,x_a) - u_{0}(t-T,x_a)
 \Bigr) \cdot 1_{+}(t-T)
\\&&{}
 +
 \Bigl(
 u(t-2T,x_b) - u_{0}(t-2T,x_b)
 \Bigr) \cdot 1_{+}(t-2T)
\\&&{}
 -
 \Bigl(
 u(t-3T,x_a) - u_{0}(t-3T,x_a)
 \Bigr) \cdot 1_{+}(t-3T)
\\&&{}
 + F_{b}(t-4T) \cdot 1_{+}(t-3T)
\\
\end{eqnarray*}
$$
 \mbox{\bf and so on. }
$$
 Take here in account that 
$$
 F_{a}(t) = 0 \,,\, F_{b}(t) = 0 \,,\, 1_{+}(t) = 0 \mbox{ falls } t < 0 \,;
$$
 in any case we see that
%
\begin{eqnarray*}
 1_{+}(t-NT) = 0 \,,\,
\\
 \mbox{ if } t < NT \,.
\end{eqnarray*}
 It follows that 
%
$$
  \mbox{ if } T \not= 0 \,, \mbox{ then }
$$
\begin{eqnarray*}
 F_{a}(t)
 &=&
 \Bigl( u(t,x_a) - u_{0}(t,x_a) \Bigr) \cdot 1_{+}(t)
\\&&{}
 -
 \Bigl(
 u(t-T,x_b) - u_{0}(t-T,x_b)
 \Bigr) \cdot 1_{+}(t-T)
\\&&{}
 +
 \Bigl(
 u(t-2T,x_a) - u_{0}(t-2T,x_a)
 \Bigr) \cdot 1_{+}(t-2T)
\\&&{}
 -
 \Bigl(
 u(t-3T,x_b) - u_{0}(t-3T,x_b)
 \Bigr) \cdot 1_{+}(t-3T)
\\&&{}
 + \cdots
\\[\smallskipamount]\\
 F_{b}(t)
 &=&
 \Bigl( u(t,x_b) - u_{0}(t,x_b) \Bigr) \cdot 1_{+}(t)
\\&&{}
 -
 \Bigl(
 u(t-T,x_a) - u_{0}(t-T,x_a)
 \Bigr) \cdot 1_{+}(t-T)
\\&&{}
 +
 \Bigl(
 u(t-2T,x_b) - u_{0}(t-2T,x_b)
 \Bigr) \cdot 1_{+}(t-2T)
\\&&{}
 -
 \Bigl(
 u(t-3T,x_a) - u_{0}(t-3T,x_a)
 \Bigr) \cdot 1_{+}(t-3T)
\\&&{}
 + \cdots
\\
\end{eqnarray*}
\newpage 
\noindent 
\fbox{
\parbox{\textwidth}{
\medskip\noindent
\begin{center}
{\large \bf
 The 
$ \cdot 1_{+}()$
 Convention. }
\end{center}
 In the following text, we will often redefine functions of 
$t$ 
 by multiplying them by 
$1_{+}(t)$, 
 i.e., by resetting, e.g.,
%
\begin{eqnarray*}
 u_{0}(t,\cdots)
 &:=&
 u_{0}(t,\cdots) \cdot 1_{+}(t)
\\
 u(t,\cdots)
 &:=&
 u(t,\cdots) \cdot 1_{+}(t)
\\
 u_{0}(t-T,\cdots)
 &:=&
 u_{0}(t-T,\cdots) \cdot 1_{+}(t-T)
\\
 &\makebox[0ex][c]{ and so on. }&
\\
\end{eqnarray*}
 Nevertheless we will sometimes write the multiplier 
$1_{+}()$,
 although we will mostly do it for emphasis,
 or there, where the formula does not become too long. 
} 
} 

\bigskip

\noindent
 To take an example of this convention implementation,
 we say that 
 the recent relations concerning 
$F_{a},  F_{b}$,  
 we may write them as 
%
%
\begin{eqnarray*}
 F_{a}(t)
 &=&
 u(t,x_a) - u_{0}(t,x_a)
 - F_{b}(t-T)
\\
 F_{b}(t)
 &=&
 u(t,x_b) - u_{0}(t,x_b)
 - F_{a}(t-T)
\\
\end{eqnarray*}
 In addition to The 
$ \cdot 1_{+}()$
 Convention, we will

\medskip\noindent 
\fbox{
\parbox{\textwidth}{
\medskip\noindent
\begin{center}
{\large \bf
 assume that 
$ T \not= 0 $ .
 }
\end{center}
} 
} 

\bigskip

\noindent

\newpage
\section%
{ Particular Cases. Recurrence Relations }
\subsection%
{The Case of 
 $ \gamma_b = 0 $}
 We begin the analysis of this case turning to the following relations, 
 which we have seen in the previous section:
%

\medskip\noindent
\begin{eqnarray*}
 u(t,x)
 &=&
 F_{a}(t-|x-x_a|/c)
 + F_{b}(t-|x-x_b|/c)
 + u_{0}(t,x)
\qquad (t \geq 0)
\\
 F_{a}(t)
 &=&
 \Bigl(
 u(t,x_a)
 - u_{0}(t,x_a)
 - F_{b}(t-T)
 \Bigr) \cdot 1_{+}(t)
\\
 F_{b}(t)
 &=&
 \Bigl(
 u(t,x_b)
 - u_{0}(t,x_b)
 - F_{a}(t-T)
 \Bigr) \cdot 1_{+}(t)
\end{eqnarray*}
  Thus, we see, that in the case of
$\gamma_b = 0$, 
 these relations become 

\medskip\noindent
\fbox{
\parbox{\textwidth}{
\begin{eqnarray*}
 u(t,x)
 &=&
 F_{a}(t-|x-x_a|/c)
 + u_{0}(t,x)
\qquad (t \geq 0)
\\
\end{eqnarray*}
\begin{eqnarray*}
 u(t,x)
 &=&
 \Bigr( u(t-|x-x_a|/c,x_a)
 - u_{0}(t-|x-x_a|/c,x_a) \Bigr) \cdot 1_{+}(t-|x-x_a|/c)
\\&&{}
 + u_{0}(t,x)
\qquad (t \geq 0)
\\
\end{eqnarray*}
 Take here into account that 
%
\begin{eqnarray*}
 u(t,x_a)
 &=&
 F_{a}(t)
 + u_{0}(t,x_a)
\qquad (t \geq 0)
\\
\end{eqnarray*}
} 
} 

\medskip\noindent
 In particular, if 
%
$$
 u(t,x_a) \equiv 0 \qquad ( t \geq 0 ) \,,
$$
 then 
%
$$
 F_{a}(t)
 = - u_{0}(t,x_a)
$$
 and, hence, 
%
\medskip\noindent
\medskip

\noindent
\fbox{
\parbox{\textwidth}{

\medskip\noindent
 if 
%
$$
 u(t,x_a) \equiv 0 \qquad ( t \geq 0 ) \,,
$$
 then 
%
\begin{eqnarray*}
 u(t,x)
 &=&
 - u_{0}(t-|x-x_a|/c,x_a) \cdot 1_{+}(t-|x-x_a|/c)
 + u_{0}(t,x)
\qquad (t \geq 0)
\\
\end{eqnarray*}
} 
} 

\medskip\noindent
\newpage\noindent

 The restriction 
$ u(t,x_a) \equiv 0 \quad ( t \geq 0 ) $
 states that the string is absolutely motionless, fixed at the point 
$x=x_a$.
 If we replace this restriction by 
%
\begin{eqnarray*}
 F_a(t)
 &:=&
 -{2\gamma_a}
 \int_0^{t}
 \Big(F_{src\ a}(\tau)\Big)d\tau \cdot 1_{+}(t) \,,
\\
 &:=&
 -{2\gamma_a}
 \int_0^{t}
 u(\tau,x_a) d\tau \cdot 1_{+}(t) \,,
\\
\end{eqnarray*}
 then we obtain: 
%
\begin{eqnarray*}
 \frac{\partial F_a(t)}{\partial t}
 &=&
 -{2\gamma_a} u(t,x_a)
\qquad\qquad\qquad (t \geq 0)
\\
 &=&
 -{2\gamma_a}( F_a(t) + u_{0}(t,x_a) )
\qquad (t \geq 0)
\\
\end{eqnarray*}
 and, finally, because of 
$F_a(0)=0$,
%
\medskip

\noindent
\fbox{
\parbox{\textwidth}{
\begin{eqnarray*}
 F_a(t)
 &=&
 -{2\gamma_a}
 \int_0^{t}
  e^{-{2\gamma_a} (t-\tau)} u_{0}(\tau,x_a) d\tau \cdot 1_{+}(t) \,,
\\[2\bigskipamount]
 u(t,x)
 &=&
  F_a(t-|x-x_a|/c,x_a)
   + u_{0}(t,x)
\qquad (t \geq 0)
\\
\end{eqnarray*}
} 
} 

\medskip\noindent

\newpage
\subsection%
{The Case of 
 $\quad u(t,x_a) = 0\,,\quad u(t,x_b) = 0 $}
 We have yet observed that in any case 
%
\begin{eqnarray*}
 F_{a}(t)
 &=&
 \Bigl( u(t,x_a) - u_{0}(t,x_a) \Bigr) \cdot 1_{+}(t)
\\&&{}
 -
 \Bigl(
 u(t-T,x_b) - u_{0}(t-T,x_b)
 \Bigr) \cdot 1_{+}(t-T)
\\&&{}
 +
 \Bigl(
 u(t-2T,x_a) - u_{0}(t-2T,x_a)
 \Bigr) \cdot 1_{+}(t-2T)
\\&&{}
 -
 \Bigl(
 u(t-3T,x_b) - u_{0}(t-3T,x_b)
 \Bigr) \cdot 1_{+}(t-3T)
\\&&{}
 + \cdots 
\\[\smallskipamount]\\
 F_{b}(t)
 &=&
 \Bigl( u(t,x_b) - u_{0}(t,x_b) \Bigr) \cdot 1_{+}(t)
\\&&{}
 -
 \Bigl(
 u(t-T,x_a) - u_{0}(t-T,x_a)
 \Bigr) \cdot 1_{+}(t-T)
\\&&{}
 +
 \Bigl(
 u(t-2T,x_b) - u_{0}(t-2T,x_b)
 \Bigr) \cdot 1_{+}(t-2T)
\\&&{}
 -
 \Bigl(
 u(t-3T,x_a) - u_{0}(t-3T,x_a)
 \Bigr) \cdot 1_{+}(t-3T)
\\&&{}
 + \cdots 
\\
\end{eqnarray*}
 Thus we infer that, if 
%
$$
 u(t,x_a) = 0\,,\quad u(t,x_b) = 0 \,,
$$
 then exactly 
%
\footnote{
 accepting 
{\bf The 
$ \cdot 1_{+}()$
 Convention },
 }
\begin{eqnarray*}
\makebox[0ex][l]{$ u(t,x) $}
\\
 &=&
 u_{0}(t,x)
\\&&{}
 -u_{0}(t-|x-x_a|/c,x_a)
 -u_{0}(t-|x-x_b|/c,x_b)
\\&&{}
 +u_{0}(t-|x-x_a|/c-T,x_b)
 +u_{0}(t-|x-x_b|/c-T,x_a)
\\&&{}
 -u_{0}(t-|x-x_a|/c-2T,x_a)
 -u_{0}(t-|x-x_b|/c-2T,x_b)
\\&&{}
 +u_{0}(t-|x-x_a|/c-3T,x_b)
 +u_{0}(t-|x-x_b|/c-3T,x_a)
\\&&{}
 \cdots
\qquad (t \geq 0)
\\
\end{eqnarray*}

\medskip

 Needs there a comment?

\newpage
\subsection%
{The Case of 
 $\quad u(t,x_a) = 0\,,\quad \gamma_b \not= \infty $ }
 It this case, we observe
\footnote{
 exploiting 
{\bf The 
$ \cdot 1_{+}()$
 Convention },
 }
 that 
%
%
%
\begin{eqnarray*}
 F_{a}(t)
 &=&
 0 - u_{0}(t,x_a)
 - F_{b}(t-T)
\\
 F_{b}(t)
 &=&
 u(t,x_b) - u_{0}(t,x_b)
 - F_{a}(t-T)
\\
\end{eqnarray*}
\begin{eqnarray*}
 F_{a}(t)
 &=&
 0 - u_{0}(t,x_a)
\\&&{}
 -
 u(t-T,x_b) + u_{0}(t-T,x_b)
\\&&{}
 + F_{a}(t-2T)
\\[\smallskipamount]\\
 F_{b}(t)
 &=&
 u(t,x_b) - u_{0}(t,x_b)
\\&&{}
 - 0 + u_{0}(t-T,x_a)
\\&&{}
 + F_{b}(t-2T)
\\
\end{eqnarray*}
\noindent 
 Thus we infer that, if 
%
\begin{eqnarray*}
 F_b(t)
 &:=&
 -{2\gamma_b}
 \int_0^{t}
 \Big(F_{src\ b}(\tau)\Big)d\tau \cdot 1_{+}(t) \,,
\\
 &:=&
 -{2\gamma_b}
 \int_0^{t}
 u(\tau,x_b) d\tau \cdot 1_{+}(t) \,,
\\
\end{eqnarray*}
 then 
%
\begin{eqnarray*}
 F_{a}(t)
 &=&
 - u_{0}(t,x_a)
\\&&{}
 -
 u(t-T,x_b) + u_{0}(t-T,x_b)
\\&&{}
 + F_{a}(t-2T)
\\[\smallskipamount]\\
 -{2\gamma_b}
 \int_0^{t}
 u(\tau,x_b) d\tau \cdot 1_{+}(t)
 &=&
 u(t,x_b) - u_{0}(t,x_b)
\\&&{}
 + u_{0}(t-T,x_a)
\\&&{}
 -
 {2\gamma_b}
 \int_0^{t-2T}
 u(\tau,x_b) d\tau \cdot 1_{+}(t-2T)
\\
\end{eqnarray*}
 We now focus on the latter relation, which we write in the following terms: 

%
%
%
\begin{eqnarray*}
 u(t,x_b)
 &=&
 -{2\gamma_b}
 \int_0^{t}
 u(\tau,x_b) d\tau
 +
 {2\gamma_b}
 \int_0^{t-2T}
 u(\tau,x_b) d\tau
\\&&{}
 + u_{0}(t,x_b) - u_{0}(t-T,x_a)
\\
\end{eqnarray*}
\newpage 
\noindent
 Next, let us define 
%
$$
 Q_{b}(t) := u(t,x_b) \,;\,
 Q_{0b}(t) := u_{0}(t,x_b) \,;\,
 Q_{0a}(t) := u_{0}(t,x_a) \,;\,
$$
 so that the former relation becomes 
%
\begin{eqnarray*}
 Q_{b}(t)
 &=&
 -{2\gamma_b}
 \int_0^{t}
 Q_{b}(\tau) d\tau
 +
 {2\gamma_b}
 \int_0^{t-2T}
 Q_{b}(\tau) d\tau
\\&&{}
 + Q_{0b}(t) - Q_{0a}(t-T)
\\
\end{eqnarray*}
\begin{eqnarray*}
 Q_{b}(t)
 &=&
 -{2\gamma_b}
 \int_{t-2T}^{t}
 Q_{b}(\tau) d\tau
\\&&{}
 + Q_{0b}(t) - Q_{0a}(t-T)
\\
\end{eqnarray*}
\begin{eqnarray*}
 \frac{\partial Q_{b}(t)}{\partial t}
 &=&
 -{2\gamma_b}
 Q_{b}(t)
 +
 {2\gamma_b}
 Q_{b}(t-2T)
\\&&{}
 + \frac{\partial ( Q_{0b}(t) - Q_{0a}(t-T) )}{\partial t}
\\
\end{eqnarray*}
 Let
%
\footnote{ Method of Variation of Constants }
%
$$
 Q_{b}(t) = e^{-{2\gamma_b}t} C(t) \,.
$$
 Then we obtain:
%
\begin{eqnarray*}
 Q_{b}(t-2T) &=& e^{-{2\gamma_b}t} e^{{2\gamma_b}T}C(t-2T) \,,
\\
 \frac{\partial Q_{b}(t)}{\partial t}
 &=&
 -{2\gamma_b} e^{-{2\gamma_b}t} C(t)
 +
 e^{-{2\gamma_b}t} \frac{\partial C(t)}{\partial t}
\\
 &=&
 -{2\gamma_b}Q_{b}(t)
 +
 e^{-{2\gamma_b}t} \frac{\partial C(t)}{\partial t}
\\
\end{eqnarray*}
\begin{eqnarray*}
 e^{-{2\gamma_b}t} \frac{\partial C(t)}{\partial t}
 &=&
 {2\gamma_b}
 Q_{b}(t-2T)
\\&&{}
 + \frac{\partial ( Q_{0b}(t) - Q_{0a}(t-T) )}{\partial t}
\\
\end{eqnarray*}
\newpage 
\noindent 
\begin{eqnarray*}
 e^{-{2\gamma_b}t} \frac{\partial C(t)}{\partial t}
 &=&
 {2\gamma_b}
 e^{-{2\gamma_b}t} e^{{2\gamma_b}T}C(t-2T)
\\&&{}
 + \frac{\partial ( Q_{0b}(t) - Q_{0a}(t-T) )}{\partial t}
\\
\end{eqnarray*}
\begin{eqnarray*}
 \frac{\partial C(t)}{\partial t}
 &=&
 {2\gamma_b} e^{{2\gamma_b}T}
 C(t-2T)
\\&&{}
 + e^{{2\gamma_b}t}
 \frac{\partial ( Q_{0b}(t) - Q_{0a}(t-T) )}{\partial t}
\\
\end{eqnarray*}
\begin{eqnarray*}
 \frac{\partial C(t)}{\partial t}
 &=&
 {2\gamma_b}
 e^{{2\gamma_b}t} Q_{b}(t-2T)
\\&&{}
 +e^{{2\gamma_b}t}
 \frac{\partial ( Q_{0b}(t) - Q_{0a}(t-T) )}{\partial t}
\\
\end{eqnarray*}
\begin{eqnarray*}
 C(t)
 &=&
 {2\gamma_b}
 \int_0^{t}
 e^{{2\gamma_b}\tau } Q_{b}(\tau-2T)
 d\tau
\\&&{}
 +
 \int_0^{t}
 e^{{2\gamma_b}\tau}
 \frac{\partial ( Q_{0b}(\tau) - Q_{0a}(\tau-T) )}{\partial \tau}
 d\tau
 +
 C(0)
\\
\end{eqnarray*}
\begin{eqnarray*}
 Q_{b}(t)
 &=&
 e^{-{2\gamma_b}t}
 {2\gamma_b}
 \int_0^{t}
 e^{{2\gamma_b}\tau } Q_{b}(\tau-2T)
 d\tau
\\&&{}
 +
 e^{-{2\gamma_b}t}
 \int_0^{t}
 e^{{2\gamma_b}\tau}
 \frac{\partial ( Q_{0b}(\tau) - Q_{0a}(\tau-T))}{\partial \tau}
 d\tau
 +
 e^{-{2\gamma_b}t} Q_{b}(0)
\\
\end{eqnarray*}
\begin{eqnarray*}
 Q_{b}(t)
 &=&
 {2\gamma_b}
 \int_0^{t}
 e^{-{2\gamma_b}(t-\tau )} Q_{b}(\tau-2T)
 d\tau
\\&&{}
 +
 \int_0^{t}
 e^{-{2\gamma_b}(t-\tau )}
 d_{\tau} 
 ( Q_{0b}(\tau) - Q_{0a}(\tau-T) )
 +
 e^{-{2\gamma_b}t} Q_{b}(0)
\\
\end{eqnarray*}
\begin{eqnarray*}
 Q_{b}(t)
 &=&
 {2\gamma_b}
 \int_0^{t}
 e^{-{2\gamma_b}(t-\tau )} Q_{b}(\tau-2T)
 d\tau
\\&&{}
 +
  I_{0b}(t) \,,
\\
\makebox[0ex][l]{ where }
\\
  I_{0b}(t)
&:=&
 \int_0^{t}
 e^{-{2\gamma_b}(t-\tau )}
 d_{\tau} 
 ( Q_{0b}(\tau) - Q_{0a}(\tau-T) )
 +
 e^{-{2\gamma_b}t} Q_{b}(0)
\\
\end{eqnarray*}

\medskip\noindent
\newpage\noindent
 and then, after returning to the terms of 
$ u \,,\, u_{0} $,
 we infer that 

\medskip\noindent
\fbox{
\parbox{\textwidth}{
\begin{eqnarray*}
 u(t,x_b)
 &=&
 {2\gamma_b}
 \int_{2T}^{t}
 e^{-{2\gamma_b}(t-\tau )} u(\tau-2T,x_b)
 d\tau
\\&&{}
 +
  I_{0b}(t) \,,
\\
\makebox[0ex][l]{ where }
\\
 I_{0b}(t)
&:=&
 \int_{0+0}^{t}
 e^{-{2\gamma_b}(t-\tau )}
 d_{\tau} 
 ( u_{0}(\tau,x_b) - u_{0}(\tau-T,x_a)1_{+}(\tau-T) )
 +
 e^{-{2\gamma_b}t} u(0,x_b)
\\
\end{eqnarray*}
} 
} 

\medskip\noindent
 Finally, recall that 

\medskip\noindent
\fbox{
\parbox{\textwidth}{
\begin{eqnarray*}
 u(t,x)
 &=&
 F_{a}(t-|x-x_a|/c)
 - F_{b}(t-|x-x_b|/c)
 + u_{0}(t,x)
\qquad (t \geq 0)
\\
\end{eqnarray*}
} 
} 

\medskip\noindent
 where
\begin{eqnarray*}
 F_{a}(t)
 &=&
 \Bigl( 0 - u_{0}(t,x_a) \Bigr) \cdot 1_{+}(t)
\\&&{}
 -
 \Bigl(
 u(t-T,x_b) - u_{0}(t-T,x_b)
 \Bigr) \cdot 1_{+}(t-T)
\\&&{}
 +
 \Bigl(
 0 - u_{0}(t-2T,x_a)
 \Bigr) \cdot 1_{+}(t-2T)
\\&&{}
 -
 \Bigl(
 u(t-3T,x_b) - u_{0}(t-3T,x_b)
 \Bigr) \cdot 1_{+}(t-3T)
\\&&{}
 + \cdots 
\\[\smallskipamount]\\
 F_{b}(t)
 &=&
 \Bigl( u(t,x_b) - u_{0}(t,x_b) \Bigr) \cdot 1_{+}(t)
\\&&{}
 -
 \Bigl(
  0 - u_{0}(t-T,x_a)
 \Bigr) \cdot 1_{+}(t-T)
\\&&{}
 +
 \Bigl(
 u(t-2T,x_b) - u_{0}(t-2T,x_b)
 \Bigr) \cdot 1_{+}(t-2T)
\\&&{}
 -
 \Bigl(
 0 - u_{0}(t-3T,x_a)
 \Bigr) \cdot 1_{+}(t-3T)
\\&&{}
 + \cdots 
\\
\end{eqnarray*}
\newpage
\subsection%
{The Case of 
 $\gamma_a \not= \infty \,,\quad \gamma_b \not= \infty $ }
%
%
 As before, we begin to analyse the situation looking at the relations 
\begin{eqnarray*}
 u(t,x) \cdot 1_{+}(t)
 &=&
 F_{a}(t-|x-x_a|/c)
 + F_{b}(t-|x-x_b|/c)
 + u_{0}(t,x) \cdot 1_{+}(t)
\\
\end{eqnarray*}
\begin{eqnarray*}
 u(t,x_a)
 &=&
 F_{a}(t)
 + F_{b}(t-T)
 + u_{0}(t,x_a)
\qquad (t \geq 0)
\\
\end{eqnarray*}
\begin{eqnarray*}
 u(t,x_b)
 &=&
 F_{a}(t-T)
 + F_{b}(t)
 + u_{0}(t,x_b)
\qquad (t \geq 0)
\\
\end{eqnarray*}
 We focus on the two latter.
%
 Putting, for short, 
%
$$
 Q_{a}(t) := u(t,x_a) \,;\,
 Q_{0a}(t) := u_{0}(t,x_a) \,;\,
 Q_{b}(t) := u(t,x_b) \,;\,
 Q_{0b}(t) := u_{0}(t,x_b) \,;\,
$$
\begin{eqnarray*}
 (V_{T}f)(t) &:=&  f(t-T) \,;\,
\\
\end{eqnarray*}
 we transform them into 
\begin{eqnarray*}
 \left(
  \begin{array}{cc}
  Q_{a} \\ Q_{b}
  \end{array}
 \right)
 &=&
 \left(
  \begin{array}{cc}
  I & V_{T} \\ V_{T} & I
  \end{array}
 \right)
 \left(
  \begin{array}{cc}
   F_{a} \\ F_{b}
  \end{array}
 \right)
 +
 \left(
  \begin{array}{cc}
  Q_{0a} \\ Q_{0b}
  \end{array}
 \right)
\end{eqnarray*}
%
 Now take into account the fact that the current situation is such that 
\begin{eqnarray*}
 F_{a}(t)
 &=&
 \displaystyle
 -{2\gamma_a}\int_0^{t} 
 u(\tau,x_a)d\tau
 =
 -{2\gamma_a}\int_0^{t} 
 Q_{a}(\tau)d\tau
\end{eqnarray*}
\begin{eqnarray*}
 F_{b}(t)
 &=&
 \displaystyle
 -{2\gamma_b}\int_0^{t} 
 u(\tau,x_b)d\tau
 =
 -{2\gamma_b}\int_0^{t} 
 Q_{b}(\tau)d\tau
\end{eqnarray*}
%
%
 As a result, 
\begin{eqnarray*}
 \frac{\partial }{\partial t}
 \left(
  \begin{array}{cc}
  F_{a} \\ F_{b}
  \end{array}
 \right)
 &=&
 \left(
  \begin{array}{cc}
  -{2\gamma_a}Q_{a} \\ -{2\gamma_b}Q_{b}
  \end{array}
 \right)
\\[\bigskipamount]
 &=&
 \left(
  \begin{array}{cc}
  -{2\gamma_a} & 0 \\ 0 & -{2\gamma_b}
  \end{array}
 \right)
 \left(
  \begin{array}{cc}
   Q_{a} \\ Q_{b}
  \end{array}
 \right)
\\[\bigskipamount]
 &=&
 \left(
  \begin{array}{cc}
  -{2\gamma_a} & 0 \\ 0 & -{2\gamma_b}
  \end{array}
 \right)
 \left(
  \begin{array}{cc}
  I & V_{T} \\ V_{T} & I
  \end{array}
 \right)
 \left(
  \begin{array}{cc}
   F_{a} \\ F_{b}
  \end{array}
 \right)
 +
 \left(
  \begin{array}{cc}
  -{2\gamma_a} & 0 \\ 0 & -{2\gamma_b}
  \end{array}
 \right)
 \left(
  \begin{array}{cc}
  Q_{0a} \\ Q_{0b}
  \end{array}
 \right)
\\[\bigskipamount]
 &=&
 \left(
  \begin{array}{cc}
  -{2\gamma_a} & 0 \\ 0 & -{2\gamma_b}
  \end{array}
 \right)
 \left(
  \begin{array}{cc}
  I & V_{T} \\ V_{T} & I
  \end{array}
 \right)
 \left(
  \begin{array}{cc}
   F_{a} \\ F_{b}
  \end{array}
 \right)
 -
 \left(
  \begin{array}{cc}
  {2\gamma_a}Q_{0a} \\ {2\gamma_b}Q_{0b}
  \end{array}
 \right)
\\[\bigskipamount]
 &=&
 \left(
  \begin{array}{cc}
  -{2\gamma_a} & -{2\gamma_a}V_{T} \\ -{2\gamma_b}V_{T} & -{2\gamma_b}
  \end{array}
 \right)
 \left(
  \begin{array}{cc}
   F_{a} \\ F_{b}
  \end{array}
 \right)
 -
 \left(
  \begin{array}{cc}
  {2\gamma_a}Q_{0a} \\ {2\gamma_b}Q_{0b}
  \end{array}
 \right)
\end{eqnarray*}
 \newpage
\noindent
 To continue the analysis, make some preparations.
%
 First, for any real 
$\gamma$,
 introduce an operation,
$E_{\gamma}$,
 defining it by
%
\begin{eqnarray*}
 \Bigl( E_{\gamma} f \Bigr)(t)
 &:=&
 e^{-2{\gamma} t}f(t) \,.
\end{eqnarray*}
 In addition, let 
$I_{0}$
 stands for the operation defined by 
\begin{eqnarray*}
 \Bigl( I_{0} f \Bigr)(t)
 &:=&
 \int_{0}^{t}f(\tau)d\tau \,.
\end{eqnarray*}
%
 Now let 
$ C_{a}, C_{b} $
 be defined by 
\footnote{ Method of Variation of Constants }
%
\begin{eqnarray*}
 \left(
  \begin{array}{cc}
  F_{a} \\ F_{b}
  \end{array}
 \right)
 &=&
 \left(
  \begin{array}{cc}
  E_{\gamma_a}C_{a} \\ E_{\gamma_b}C_{b}
  \end{array}
 \right)
\end{eqnarray*}
 Then 
\begin{eqnarray*}
 \frac{\partial }{\partial t}
 \left(
  \begin{array}{cc}
  F_{a} \\ F_{b}
  \end{array}
 \right)
 &=&
 \left(
  \begin{array}{cc}
  -2{\gamma_a}E_{\gamma_a}C_{a} \\ -2{\gamma_b}E_{\gamma_b}C_{b}
  \end{array}
 \right)
 +
 \left(
  \begin{array}{cc}
  E_{\gamma_a}\frac{\partial }{\partial t}C_{a}
  \\
  E_{\gamma_b}\frac{\partial }{\partial t}C_{b}
  \end{array}
 \right)
\\[\bigskipamount]
 &=&
 \left(
  \begin{array}{cc}
  -2{\gamma_a} & 0  \\ 0 & {\gamma_b}
  \end{array}
 \right)
 \left(
  \begin{array}{cc}
  F_{a} \\ F_{b}
  \end{array}
 \right)
 +
 \left(
  \begin{array}{cc}
  E_{\gamma_a} & 0 \\ 0 & E_{\gamma_b}
  \end{array}
 \right)
\frac{\partial }{\partial t}
 \left(
  \begin{array}{cc}
  C_{a} \\ C_{b}
  \end{array}
 \right)
\\
\end{eqnarray*}
 and therefore 
\begin{eqnarray*}
 \left(
  \begin{array}{cc}
  E_{\gamma_a} & 0 \\ 0 & E_{\gamma_b}
  \end{array}
 \right)
\frac{\partial }{\partial t}
 \left(
  \begin{array}{cc}
  C_{a} \\ C_{b}
  \end{array}
 \right)
 &=&
 \left(
  \begin{array}{cc}
  0 & -2{\gamma_a}V_{T} \\ -2{\gamma_b}V_{T} & 0
  \end{array}
 \right)
 \left(
  \begin{array}{cc}
  F_{a} \\ F_{b}
  \end{array}
 \right)
 -
 \left(
  \begin{array}{cc}
  2{\gamma_a}Q_{0a} \\ 2{\gamma_b}Q_{0b}
  \end{array}
 \right)
\\
\end{eqnarray*}
\begin{eqnarray*}
\frac{\partial }{\partial t}
 \left(
  \begin{array}{cc}
  C_{a} \\ C_{b}
  \end{array}
 \right)
 &=&
 \left(
  \begin{array}{cc}
  0 & -2{\gamma_a}E_{\gamma_a}^{-1}V_{T}
 \\
  -2{\gamma_b}E_{\gamma_b}^{-1}V_{T} & 0
  \end{array}
 \right)
 \left(
  \begin{array}{cc}
  F_{a} \\ F_{b}
  \end{array}
 \right)
 -
 \left(
  \begin{array}{cc}
  2{\gamma_a}E_{\gamma_a}^{-1}Q_{0a}
 \\
  2{\gamma_b}E_{\gamma_b}^{-1}Q_{0b}
  \end{array}
 \right)
\\
\end{eqnarray*}
 Note that 
$ F_{a}(0) = 0$ ,
$ F_{b}(0) = 0$ ,
$ C_{a}(0) = 0$ ,
$ C_{b}(0) = 0$ .
 Hence 
\begin{eqnarray*}
 \left(
  \begin{array}{cc}
  C_{a} \\ C_{b}
  \end{array}
 \right)
 &=&
 \left(
  \begin{array}{cc}
  0 & -2{\gamma_a}I_{0}E_{\gamma_a}^{-1}V_{T}
 \\
  -2{\gamma_b}I_{0}E_{\gamma_b}^{-1}V_{T} & 0
  \end{array}
 \right)
 \left(
  \begin{array}{cc}
  F_{a} \\ F_{b}
  \end{array}
 \right)
 -
 \left(
  \begin{array}{cc}
  2{\gamma_a}I_{0}E_{\gamma_a}^{-1}Q_{0a}
 \\
  2{\gamma_b}I_{0}E_{\gamma_b}^{-1}Q_{0b}
  \end{array}
 \right)
\\
\end{eqnarray*}
 and then 
\begin{eqnarray*}
 \left(
  \begin{array}{cc}
  F_{a} \\ F_{b}
  \end{array}
 \right)
 &=&
 \left(
  \begin{array}{cc}
  0 & -2{\gamma_a}E_{\gamma_a}I_{0}E_{\gamma_a}^{-1}V_{T}
 \\
  -2{\gamma_b}E_{\gamma_b}I_{0}E_{\gamma_b}^{-1}V_{T} & 0
  \end{array}
 \right)
 \left(
  \begin{array}{cc}
  F_{a} \\ F_{b}
  \end{array}
 \right)
 \\&&{}
 -
 \left(
  \begin{array}{cc}
  2{\gamma_a}E_{\gamma_a}I_{0}E_{\gamma_a}^{-1}Q_{0a}
 \\
  2{\gamma_b}E_{\gamma_b}I_{0}E_{\gamma_b}^{-1}Q_{0b}
  \end{array}
 \right)
\\
\end{eqnarray*}
 i.e.,
\begin{eqnarray*}
  F_{a}(t)
 &=&
  -2{\gamma_a}
 \int_{0}^{t}e^{-2{\gamma_a}(t-\tau)}
  F_{b}(\tau-T)d\tau
  -2{\gamma_a}
 \int_{0}^{t}e^{-2{\gamma_a}(t-\tau)}
 Q_{0a}(\tau)d\tau
 \\
  F_{b}(t)
 &=&
  -2{\gamma_b}
 \int_{0}^{t}e^{-2{\gamma_b}(t-\tau)}
  F_{a}(\tau-T)d\tau
  -2{\gamma_b}
 \int_{0}^{t}e^{-2{\gamma_b}(t-\tau)}
 Q_{0b}(\tau)d\tau
 \\
\end{eqnarray*}
 equivalently, 
\begin{eqnarray*}
  F_{a}(t)
 &=&
  -2{\gamma_a}
 \int_{0}^{t}e^{-2{\gamma_a}(t-\tau)}
  F_{b}(\tau-T)d\tau
  -2{\gamma_a}
 \int_{0}^{t}e^{-2{\gamma_a}(t-\tau)}
 u_{0}(\tau,x_a)d\tau
 \\
  F_{b}(t)
 &=&
  -2{\gamma_b}
 \int_{0}^{t}e^{-2{\gamma_b}(t-\tau)}
  F_{a}(\tau-T)d\tau
  -2{\gamma_b}
 \int_{0}^{t}e^{-2{\gamma_b}(t-\tau)}
 u_{0}(\tau,x_b)d\tau
 \\
\end{eqnarray*}
\footnote{ Note that 
$ F_{a}(0) = 0$ ,
$ F_{b}(0) = 0$ ,
$ C_{a}(0) = 0$ ,
$ C_{b}(0) = 0$ .
} 
\par\medskip\noindent
 Finally, recall that 

\medskip\noindent
\fbox{
\parbox{\textwidth}{
\begin{eqnarray*}
 u(t,x)
 &=&
 F_{a}(t-|x-x_a|/c)
 + F_{b}(t-|x-x_b|/c)
 + u_{0}(t,x)
\qquad (t \geq 0)
\\
\end{eqnarray*}
} 
} 

\medskip\noindent

\newpage\noindent 
\section%
{ Explicit Relations }
\subsection%
{ The Algebra of
 ${\bf I}_{\gamma, T_0}$
 }
 In the previous section we have seen the recurrence relations, 
 which normally contain "retarded integral operations".
 When solving such relations, 
 a usual machinery involves a treatment of the various compositions
 of the operations that build the relations. 
 So,
 we try to find an effective representation
 of the referred compositions.

 Now, let 
${\bf I}_{\gamma, T_0}$
 denote the operation defined by
\footnote
{
 recall that, in the context, 
$\int_a^b$
 means 
$\int_a^b \cdot 1_{+}(b-a) $
} 
\begin{eqnarray*}
 \Bigl( {\bf I}_{\gamma, T_0} f \Bigr)(t)
 &:=&
 \int_{T_0}^{t}
 e^{-{2\gamma}(t-\tau)} f(\tau-T_0)
 d\tau 
\\&&{} 
 \equiv
 \int_{T_0}^{t}
 e^{-{2\gamma}(t-\tau)} f(\tau-T_0)
 d\tau \cdot 1_{+}(t-T_0)
\\&&{} 
 =
 \int_{0}^{t-T_0}
 e^{-{2\gamma}(t-T_0-\tau)} f(\tau)
 d\tau \cdot 1_{+}(t-T_0)
\\&&{}
 \equiv
 \int_{0}^{t-T_0}
 e^{-{2\gamma}(t-T_0-\tau)} f(\tau)
 d\tau \cdot
\end{eqnarray*}
 Firstly, notice that
\begin{eqnarray*}
 \Bigl( {\bf I}_{\gamma, T_0} f \Bigr)(t)
 &=&
 \int_{0}^{t}
 e^{-{2\gamma}(t-T_0-\tau)}
 f(\tau) \cdot 1_{+}(t-T_0-\tau)
 d\tau \cdot 1_{+}(t)
\end{eqnarray*}
 In particular, the operations under consideration
 are usual convolution operations, and as such they are commutative:
\begin{eqnarray*}
  {\bf I}_{\gamma_1, T_1}{\bf I}_{\gamma_2, T_2}
 &=&
  {\bf I}_{\gamma_2, T_2}{\bf I}_{\gamma_1, T_1} \,. 
\end{eqnarray*}
 In addition notice that 
\begin{eqnarray*}
 \Bigl( {\bf I}_{\gamma, T_0} f \Bigr)(t+T_0)
 &=&
 \int_{0}^{t}
 e^{-{2\gamma}(t-\tau)} f(\tau)
 d\tau \cdot 1_{+}(t)
\\
 &=&
 \Bigl( {\bf I}_{\gamma,0} f \Bigr)(t)
\end{eqnarray*}
 Next, let us calculate
$$
  {\bf I}_{\gamma, T_1}{\bf I}_{\gamma, T_2}
$$
 The usual way is:
\begin{eqnarray*}
\makebox[5ex][l]{$\displaystyle
 \Bigl( {\bf I}_{\gamma, T_1}{\bf I}_{\gamma, T_2} f \Bigr)(t)
$} 
\\
&&{} =
 \int_{0}^{t-T_1}
 e^{-{2\gamma}(t-T_1-\tau)}
 \Bigl( {\bf I}_{\gamma, T_2} f \Bigr)(\tau)
 d\tau \cdot 1_{+}(t-T_1)
\\&&{} =
 \int_{0}^{t-T_1}
 e^{-{2\gamma}(t-T_1-\tau)}
 \Bigl( {\bf I}_{\gamma, T_2} f \Bigr)(\tau)
 \cdot 1_{+}(\tau-T_2)
 d\tau \cdot 1_{+}(t-T_1)
\\&&{} =
 \int_{0}^{t-T_1}
 e^{-{2\gamma}(t-T_1-\tau)}
 \Bigl( {\bf I}_{\gamma, T_2} f \Bigr)(\tau)
 \cdot 1_{+}(\tau-T_2)
 d\tau \cdot 1_{+}(t-T_1-T_2)
\\&&{} =
 \int_{T_2}^{t-T_1}
 e^{-{2\gamma}(t-T_1-\tau)}
 \Bigl( {\bf I}_{\gamma, T_2} f \Bigr)(\tau)
 d\tau \cdot 1_{+}(t-T_1-T_2)
\\&&{} =
 \int_{0}^{t-T_1-T_2}
 e^{-{2\gamma}(t-T_1-T_2-\tau)}
 \Bigl( {\bf I}_{\gamma, T_2} f \Bigr)(\tau+T_2)
 d\tau \cdot 1_{+}(t-T_1-T_2)
\\&&{} =
 \int_{0}^{t-T_1-T_2}
 e^{-{2\gamma}(t-T_1-T_2-\tau)}
 \Bigl( {\bf I}_{\gamma, 0} f \Bigr)(\tau)
 d\tau \cdot 1_{+}(t-T_1-T_2)
\\&&{} =
 \Bigl( {\bf I}_{\gamma, T_1+T_2}{\bf I}_{\gamma, 0} f \Bigr)(t)
\end{eqnarray*}
 Thus, we have seen that 
\begin{eqnarray*}
 {\bf I}_{\gamma, T_1}{\bf I}_{\gamma, T_2}
 &=&
 {\bf I}_{\gamma, T_1+T_2}{\bf I}_{\gamma, 0} \,.
\end{eqnarray*}
  Consequences which we need:
\begin{eqnarray*}
 {\bf I}_{\gamma_b, 2T}^N
 &=&
 {\bf I}_{\gamma_b, 2NT}{\bf I}_{\gamma_b, 0}^{N-1} \,,
\end{eqnarray*}
\begin{eqnarray*}
 ({\bf I}_{\gamma_0, T})^{2N}
 &=&
 {\bf I}_{\gamma_0, 2NT}{\bf I}_{\gamma_0, 0}^{2N-1} \,,
\end{eqnarray*}
\begin{eqnarray*}
 ({\bf I}_{\gamma_0, T})^{2N+1}
 &=&
 {\bf I}_{\gamma_0, (2N+1)T}{\bf I}_{\gamma_0, 0}^{2N} \,,
\end{eqnarray*}
\begin{eqnarray*}
 ({\bf I}_{\gamma_a, T}{\bf I}_{\gamma_b, T})^N
 &=&
 ({\bf I}_{\gamma_b, T}{\bf I}_{\gamma_a, T})^N
\\
 &=&
 {\bf I}_{\gamma_b, NT}{\bf I}_{\gamma_a, NT}
 {\bf I}_{\gamma_b, 0}^{N-1}{\bf I}_{\gamma_a, 0}^{N-1} \,.
\end{eqnarray*}
 In addition, let 
$E_{\gamma}$
 denote the operation defined by 
\begin{eqnarray*}
 \Bigl( E_{\gamma} f \Bigr)(t)
 &:=&
 e^{\gamma t}f(t) \,.
\end{eqnarray*}
 Then we can write
\begin{eqnarray*}
 {\bf I}_{\gamma, 0} 
 &=&
 E_{\gamma}^{-1}{\bf I}_{0,0}E_{\gamma} \,,
\end{eqnarray*}
 and then 
\begin{eqnarray*}
 {\bf I}_{\gamma, 0}^2 
 &=&
 E_{\gamma}^{-1}{\bf I}_{0,0}E_{\gamma}E_{\gamma}^{-1}{\bf I}_{0,0}E_{\gamma}
\\
&&{} =
E_{\gamma}^{-1}{\bf I}_{0,0}^2E_{\gamma} \,.
\end{eqnarray*}
 Iterating this kind of arguments we infer that 
\begin{eqnarray*}
 {\bf I}_{\gamma, T_0}^N
 &=&
 {\bf I}_{\gamma, NT_0}E_{\gamma}^{-1}{\bf I}_{0,0}^{N-1}E_{\gamma} \,,
\\
 {\bf I}_{\gamma, T_0}^N {\bf I}_{\gamma, 0}
 &=&
 {\bf I}_{\gamma, NT_0}E_{\gamma}^{-1}{\bf I}_{0,0}^{N}E_{\gamma}
\end{eqnarray*}
 and then 
\begin{eqnarray*}
 ( {\bf I}_{\gamma, T_0}^N f )(t)
 &=&
 \int_{0}^{t-NT_0}
 \int_{0}^{\tau}
 \int_{0}^{\tau_1}
 \cdots
 \int_{0}^{\tau_{N-2}}
 e^{-{2\gamma}(t-NT_0-\tau_{N-2})} f(\tau_{N-2})
 d\tau_{N-2}
 \cdots
 d\tau_1
 d\tau
\\
 &=&
 \int_{0}^{t-NT_0}
 \frac{(t-NT_0-\tau)^{N-1}}{(N-1)!}
 e^{-{2\gamma}(t-NT_0-\tau)} f(\tau)
 d\tau
\\
 &\equiv&
 \int_{0}^{t}
 \frac{(t-NT_0-\tau)^{N-1}}{(N-1)!}
 e^{-{2\gamma}(t-NT_0-\tau)} f(\tau) \cdot 1_{+}(t-NT_0-\tau)
 d\tau \,,
\end{eqnarray*}
\begin{eqnarray*}
 ( {\bf I}_{\gamma, T_0}^N {\bf I}_{\gamma, 0} f )(t)
 &=&
 \int_{0}^{t-NT_0}
 \frac{(t-NT_0-\tau)^{N}}{N!}
 e^{-{2\gamma}(t-NT_0-\tau)} f(\tau)
 d\tau
\\
 &\equiv&
 \int_{0}^{t}
 \frac{(t-NT_0-\tau)^{N}}{N!}
 e^{-{2\gamma}(t-NT_0-\tau)} f(\tau) \cdot 1_{+}(t-NT_0-\tau)
 d\tau \,.
\end{eqnarray*}
 Now, we go on to see consequences of these consequences.
\newpage\noindent 
\subsection%
{ Explicit Relations:
 The Case of 
 $\quad u(t,x_a) = 0\,,\quad \gamma_b \not= \infty $ }
 We have seen that

\begin{eqnarray*}
 u(t,x_b)
 &=&
 {2\gamma_b}
 \int_{2T}^{t}
 e^{-{2\gamma_b}(t-\tau )} u(\tau-2T,x_b)
 d\tau
\\&&{}
 +
  I_{0b}(t) \,,
\\
\makebox[0ex][l]{ where }
\\
 I_{0b}(t)
&:=&
 \int_{0+0}^{t}
 e^{-{2\gamma_b}(t-\tau )}
 d_{\tau} 
 ( u_{0}(\tau,x_b) - u_{0}(\tau-T,x_a)1_{+}(\tau-T) )
 +
 e^{-{2\gamma_b}t} u(0,x_b)
 \,.
\\
\end{eqnarray*}
 Recall, 
${\bf I}_{\gamma, T_0}$
 denotes the operation defined by
\begin{eqnarray*}
 \Bigl( {\bf I}_{\gamma, T_0} f \Bigr)(t)
 &:=&
 \int_{T_0}^{t}
 e^{-{2\gamma}(t-\tau)} f(\tau-T_0)
 d\tau 
\\&&{} 
 =
 e^{2{\gamma}T_0}
 \int_{0}^{t-T_0}
 e^{-{2\gamma}(t-\tau)} f(\tau)
 d\tau 
\\&&{}
 \equiv
 e^{2{\gamma}T_0}
 \int_{-\infty}^{\infty}
 e^{-{2\gamma_b}(t-\tau)} f(\tau) \cdot 1_{+}(t-T_0-\tau) \cdot 1_{+}(\tau)
 d\tau \cdot 1_{+}(t-T_0) \,.
\end{eqnarray*}
 Thus, we can write 
\begin{eqnarray*}
 u(t,x_b)
 &=&
 I_{0b}(t)
\,,\quad ( \mbox{ if } 0 \leq t < 2T )
\end{eqnarray*}
\begin{eqnarray*}
 u(t,x_b)
 &=&
 \biggl(\Bigl(
 1 + ({2\gamma_b}) {\bf I}_{\gamma_b, 2T}
 \Bigr) I_{0b}\biggr) (t)
\,,\quad ( \mbox{ if } 0 \leq t < 4T )
\end{eqnarray*}
\begin{eqnarray*}
 u(t,x_b)
 &=&
 \biggl(\Bigl(
 1 + ({2\gamma_b}) {\bf I}_{\gamma_b, 2T}
 + ({2\gamma_b})^2 {\bf I}_{\gamma, 2T}^2
 + \cdots
 + ({2\gamma_b})^N {\bf I}_{\gamma_b, 2T}^N
 \Bigr) I_{0b}\biggr) (t)
\,,\quad ( \mbox{ if } 0 \leq t < (N+1)2T )
\end{eqnarray*}
 So, iterating, 
\begin{eqnarray*}
 u(t,x_b)
 &=&
 \biggl(\Bigl(
 1 + ({2\gamma_b}) {\bf I}_{\gamma_b, 2T}
 + ({2\gamma_b})^2 {\bf I}_{\gamma, 2T}^2
 + \cdots
 + ({2\gamma_b})^N {\bf I}_{\gamma_b, 2T}^N \cdots 
 \Bigr) I_{0b}\biggr) (t)
\end{eqnarray*}
 Finally, by recalling that 
\begin{eqnarray*}
 ( {\bf I}_{\gamma, T_0}^N f )(t)
 &=&
 \int_{0}^{t}
 \frac{(t-NT_0-\tau)^{N-1}}{(N-1)!}
 e^{-{2\gamma}(t-NT_0-\tau)} f(\tau) \cdot 1_{+}(t-NT_0-\tau)
 d\tau
\end{eqnarray*}
 and after introducing 
\begin{eqnarray*}
\makebox[3ex][l]{$ Exp(\lambda,T_0,t) $}
\\&:=&
 1 \cdot 1_{+}(t) + \lambda (t-T_0) \cdot 1_{+}(t-T_0)
 + \lambda^2 \frac{(t-2T_0)^2}{2!} \cdot 1_{+}(t-2T_0) 
\\&&{}
 + \cdots + \lambda^N \frac{(t-NT_0)^N}{N!} \cdot 1_{+}(t-NT_0) \cdots
\end{eqnarray*}
 we conclude that
\begin{eqnarray*}
 u(t,x_b)
 &=&
  I_{0b}(t)
\\&&{}
 +
 ({2\gamma_b})
 \int_{2T}^{t}
 e^{-{2\gamma_b}(t-\tau )}
  I_{0b}(\tau-2T)
 d\tau
\\&&{}
 +
 ({2\gamma_b})^2
 \int_{4T}^{t}
 \int_{4T}^{\tau}
 e^{-{2\gamma_b}(t-\tau_1 )} I_{0b}(\tau_1-4T)
 d\tau_1
 d\tau
\\&&{}
 + \cdots 
\\
 &=&
  I_{0b}(t)
\\&&{}
 +
 ({2\gamma_b})
 \int_{0}^{t-2T}
 e^{-{2\gamma_b}(t-2T-\tau )}
  I_{0b}(\tau)
 d\tau
\\&&{}
 +
 ({2\gamma_b})^2
 \int_{0}^{t-4T}
 \int_{0}^{\tau}
 e^{-{2\gamma_b}(t-4T-\tau_1 )} I_{0b}(\tau_1)
 d\tau_1
 d\tau
\\&&{}
 + \cdots 
\\
\\&=&
 I_{0b}(t)
 +2{\gamma_b}e^{4{\gamma_b}T}
 \int_{0}^{t}
 Exp(2{\gamma_b}e^{4{\gamma_b}T}, 2T, t-\tau-2T)
 e^{-{2\gamma_b}(t-\tau )}
  I_{0b}(\tau)
 d\tau
\\
\\
\makebox[0ex][l]{ where }
\\
 I_{0b}(t)
 &:=&
 \int_{0+0}^{t}
 e^{-{2\gamma_b}(t-\tau )}
 d_{\tau} 
 ( u_{0}(\tau,x_b) - u_{0}(\tau-T,x_a)1_{+}(\tau-T) )
 +
 e^{-{2\gamma_b}t} u(0,x_b) \,.
\\
\end{eqnarray*}

\newpage\noindent 
\subsection%
{ Explicit Relations:
 The Case of 
 $\gamma_a \not= \infty \,,\quad \gamma_b \not= \infty $ }
 Imitating the arguments of the previous subsections, we observe that
\begin{eqnarray*}
 \left(
  \begin{array}{cc}
  F_{a} \\ F_{b}
  \end{array}
 \right)
 &=&
\left(
 \begin{array}{cc}
 0 & -2{\gamma_a}{\bf I}_{\gamma_a, T}
 \\
  -2{\gamma_b}{\bf I}_{\gamma_b, T} & 0
 \end{array}
\right)
 \left(
  \begin{array}{cc}
  F_{a} \\ F_{b}
  \end{array}
 \right)
 -
\left(
 \begin{array}{c}
   2{\gamma_a}{\bf I}_{\gamma_a, 0}Q_{0a}
 \\
   2{\gamma_b}{\bf I}_{\gamma_b, 0}Q_{0b} 
 \end{array}
\right)
\end{eqnarray*}
 So, we infer that 
\begin{eqnarray*}
 \left(
  \begin{array}{cc}
  F_{a} \\ F_{b}
  \end{array}
 \right)
 &=&
 -
\sum_{N=0}^{\infty}
\left(
 \begin{array}{cc}
 0 & -2{\gamma_a}{\bf I}_{\gamma_a, T}
 \\
  -2{\gamma_b}{\bf I}_{\gamma_b, T} & 0
 \end{array}
\right)^N
\left(
 \begin{array}{c}
   2{\gamma_a}{\bf I}_{\gamma_a, 0}Q_{0a}
 \\
   2{\gamma_b}{\bf I}_{\gamma_b, 0}Q_{0b} 
 \end{array}
\right)
\end{eqnarray*}
 Since 
\begin{eqnarray*}
 {\bf I}_{\gamma_b, T}{\bf I}_{\gamma_a, T}
&=&
 {\bf I}_{\gamma_a, T}{\bf I}_{\gamma_b, T} \,,
\end{eqnarray*}
\begin{eqnarray*}
\left(
 \begin{array}{cc}
 0 & -2{\gamma_a}{\bf I}_{\gamma_a, T}
 \\
  -2{\gamma_b}{\bf I}_{\gamma_b, T} & 0
 \end{array}
\right)^{2n}
 &=&
 (4{\gamma_a}{\gamma_b})^n ({\bf I}_{\gamma_a, T}{\bf I}_{\gamma_b, T})^n
\left(
 \begin{array}{cc}
 1 & 0
 \\
 0 & 1
 \end{array}
\right) \,,
\end{eqnarray*}
\begin{eqnarray*}
\left(
 \begin{array}{cc}
 0 & -2{\gamma_a}{\bf I}_{\gamma_a, T}
 \\
  -2{\gamma_b}{\bf I}_{\gamma_b, T} & 0
 \end{array}
\right)^{2n+1}
 &=&
 -
 (4{\gamma_a}{\gamma_b})^n ({\bf I}_{\gamma_a, T}{\bf I}_{\gamma_b, T})^n
\left(
 \begin{array}{cc}
 0 & 2\gamma_a{\bf I}_{\gamma_a, T}
 \\
 2\gamma_b {\bf I}_{\gamma_b, T} & 0
 \end{array}
\right) \,,
\end{eqnarray*}
 we infer that 
\begin{eqnarray*}
 \left(
  \begin{array}{cc}
  F_{a} \\ F_{b}
  \end{array}
 \right)
 &=&
 -
\sum_{n=0}^{\infty}
 (4{\gamma_a}{\gamma_b})^n ({\bf I}_{\gamma_a, T}{\bf I}_{\gamma_b, T})^n
\left(
 \begin{array}{cc}
 1 & 0
 \\
 0 & 1
 \end{array}
\right)
\left(
 \begin{array}{c}
   2{\gamma_a}{\bf I}_{\gamma_a, 0}Q_{0a}
 \\
   2{\gamma_b}{\bf I}_{\gamma_b, 0}Q_{0b} 
 \end{array}
\right)
\\
&&{} +
\sum_{n=0}^{\infty}
 (4{\gamma_a}{\gamma_b})^n ({\bf I}_{\gamma_a, T}{\bf I}_{\gamma_b, T})^n
\left(
 \begin{array}{cc}
 0 & 2\gamma_a{\bf I}_{\gamma_a, T}
 \\
 2\gamma_b {\bf I}_{\gamma_b, T} & 0
 \end{array}
\right)
\left(
 \begin{array}{c}
   2{\gamma_a}{\bf I}_{\gamma_a, 0}Q_{0a}
 \\
   2{\gamma_b}{\bf I}_{\gamma_b, 0}Q_{0b} 
 \end{array}
\right)
\\[\bigskipamount]\\
 &=&
 -
\sum_{n=0}^{\infty}
 (4{\gamma_a}{\gamma_b})^n ({\bf I}_{\gamma_a, T}{\bf I}_{\gamma_b, T})^n
\left(
 \begin{array}{c}
   2{\gamma_a}{\bf I}_{\gamma_a, 0}Q_{0a}
 \\
   2{\gamma_b}{\bf I}_{\gamma_b, 0}Q_{0b} 
 \end{array}
\right)
\\
&&{} +
\sum_{n=0}^{\infty}
(4{\gamma_a}{\gamma_b})^n ({\bf I}_{\gamma_a, T}{\bf I}_{\gamma_b, T})^n
\left(
 \begin{array}{c}
   2{\gamma_a}{\bf I}_{\gamma_a, T}2{\gamma_b}{\bf I}_{\gamma_b, 0}Q_{0b}
 \\
   2{\gamma_b}{\bf I}_{\gamma_b, T}2{\gamma_a}{\bf I}_{\gamma_a, 0}Q_{0a} 
 \end{array}
\right)
\end{eqnarray*}
 Unfortunately, I do not know a good representation of this formula, 
 in the general case. Nevertheless, if
$$
  \gamma_a = \gamma_b =: \gamma_0  \,,
$$
 then the situation becomes very similar to that of the previous subsection:
 we infer that 
\begin{eqnarray*}
 \left(
  \begin{array}{cc}
  F_{a} \\ F_{b}
  \end{array}
 \right)
 &=&
 -
\sum_{n=0}^{\infty}
 (2{\gamma_0})^{2n}
\left(
 \begin{array}{cc}
 1 & 0
 \\
 0 & 1
 \end{array}
\right)
 {\bf I}_{\gamma_0, T}^{2n}
\left(
 \begin{array}{c}
   2{\gamma_0}{\bf I}_{\gamma_0, 0}Q_{0a}
 \\
   2{\gamma_0}{\bf I}_{\gamma_0, 0}Q_{0b} 
 \end{array}
\right)
\\
&&{} +
\sum_{n=0}^{\infty}
 (2{\gamma_0})^{2n+1}
\left(
 \begin{array}{cc}
 0 & 1
 \\
 1 & 0
 \end{array}
\right)
 {\bf I}_{\gamma_0, T}^{2n+1}
\left(
 \begin{array}{c}
   2{\gamma_0}{\bf I}_{\gamma_0, 0}Q_{0b}
 \\
   2{\gamma_0}{\bf I}_{\gamma_0, 0}Q_{0a} 
 \end{array}
\right)
\\[\bigskipamount]\\
 &=&
 -
\sum_{n=0}^{\infty}
 (2{\gamma_0})^{2n}
 {\bf I}_{\gamma_0, T}^{2n}
\left(
 \begin{array}{c}
   2{\gamma_0}{\bf I}_{\gamma_0, 0}Q_{0a}
 \\
   2{\gamma_0}{\bf I}_{\gamma_0, 0}Q_{0b} 
 \end{array}
\right)
\\
&&{} +
\sum_{n=0}^{\infty}
 (2{\gamma_0})^{2n+1}
 {\bf I}_{\gamma_0, T}^{2n+1}
\left(
 \begin{array}{c}
   2{\gamma_0}{\bf I}_{\gamma_0, 0}Q_{0b}
 \\
   2{\gamma_0}{\bf I}_{\gamma_0, 0}Q_{0a} 
 \end{array}
\right)
\end{eqnarray*}
 Finally,
 as in the previous subsection,
 by recalling that 
\begin{eqnarray*}
 ( {\bf I}_{\gamma, T_0}^{2n} {\bf I}_{\gamma, 0} f )(t)
 &=&
 \int_{0}^{t}
 \frac{(t-2nT_0-\tau)^{2n}}{(2n)!}
 e^{-{2\gamma}(t-2nT_0-\tau)} f(\tau) \cdot 1_{+}(t-2nT_0-\tau)
 d\tau
\end{eqnarray*}
\begin{eqnarray*}
 ( {\bf I}_{\gamma, T_0}^{2n+1} {\bf I}_{\gamma, 0} f )(t)
 &=&
 \int_{0}^{t}
 \frac{(t-(2n+1)T_0-\tau)^{2n+1}}{(2n+1)!}
 e^{-{2\gamma}(t-(2n+1)T_0-\tau)} f(\tau) \cdot 1_{+}(t-(2n+1)T_0-\tau)
 d\tau
\end{eqnarray*}
 and after introducing 
\begin{eqnarray*}
\makebox[3ex][l]{$ Sinh(\lambda,T_0,t) $}
\\&:=&
 \lambda (t-T_0) \cdot 1_{+}(t-T_0)
 + \lambda^3 \frac{(t-3T_0)^3}{3!} \cdot 1_{+}(t-3T_0) 
\\&&{}
 + \cdots + \lambda^{2n+1} \frac{(t-(2n+1)T_0)^{2n+1}}{(2n+1)!}
 \cdot 1_{+}(t-(2n+1)T_0) \cdots
\\
\makebox[3ex][l]{$ Cosh(\lambda,T_0,t) $}
\\&:=&
 1 \cdot 1_{+}(t)
 + \lambda^2 \frac{(t-2T_0)^2}{2!} \cdot 1_{+}(t-2T_0) 
\\&&{}
 + \cdots + \lambda^{2n} \frac{(t-2nT_0)^{2n}}{(2n)!}
 \cdot 1_{+}(t-2nT_0) \cdots
\end{eqnarray*}
 we observe that
\begin{eqnarray*}
 \Bigl(
\sum_{n=0}^{\infty}
 (2{\gamma_0})^{2n}
 {\bf I}_{\gamma_0, T}^{2n} {\bf I}_{\gamma_0, 0} f
 \Bigr)(t)
&=&
 \int_{0}^{t}Cosh(2{\gamma_0}e^{2{\gamma_0}T} ,T ,t-\tau)
 e^{-{2\gamma_0}(t-\tau)} f(\tau) d\tau
\end{eqnarray*}
\begin{eqnarray*}
 \Bigl(
\sum_{n=0}^{\infty}
 (2{\gamma_0})^{2n+1}
 {\bf I}_{\gamma_0, T}^{2n+1} {\bf I}_{\gamma_0, 0} f
 \Bigr)(t)
&=&
 \int_{0}^{t}Sinh(2{\gamma_0}e^{2{\gamma_0}T} ,T ,t-\tau)
  e^{-{2\gamma_0}(t-\tau)} f(\tau) d\tau
\end{eqnarray*}

\bigskip\noindent
 The conclusion is evident.

\noindent
\begin{Remark}{.}\rm
\begin{eqnarray*}
 ( {\bf I}_{\gamma_0, T}^{N} {\bf I}_{\gamma_0, 0} f )(t)
 &=&
 \int_{0}^{t-NT}
 \frac{(t-NT-\tau)^{N}}{N!}
 e^{-{2\gamma_0}(t-NT-\tau)} f(\tau)
 d\tau
\\
 &=&
 \int_{NT}^{t}
 \frac{(t-\tau)^{N}}{N!}
 e^{-{2\gamma_0}(t-\tau)} f(\tau-NT)
 d\tau
\end{eqnarray*}
 Hence 
\begin{eqnarray*}
 \Bigl(
\sum_{n=0}^{\infty}
 (2{\gamma_0})^{2n}
 {\bf I}_{\gamma_0, T}^{2n} {\bf I}_{\gamma_0, 0} f
 \Bigr)(t)
&=&
\sum_{n=0}^{\infty}
 (2{\gamma_0})^{2n}
 \int_{2nT}^{t}
 \frac{(t-\tau)^{2n}}{(2n)!}
 e^{-{2\gamma_0}(t-\tau)} f(\tau-2nT) d\tau
\end{eqnarray*}
\begin{eqnarray*}
 \Bigl(
\sum_{n=0}^{\infty}
 (2{\gamma_0})^{2n+1}
 {\bf I}_{\gamma_0, T}^{2n+1} {\bf I}_{\gamma_0, 0} f
 \Bigr)(t)
&=&
\sum_{n=0}^{\infty}
 (2{\gamma_0})^{2n+1}
 \int_{(2n+1)T}^{t}
 \frac{(t-\tau)^{2n+1}}{(2n+1)!}
  e^{-{2\gamma_0}(t-\tau)} f(\tau-(2n+1)T) d\tau
\end{eqnarray*}
$\Box$
\end{Remark}

\newpage 

\bibliographystyle{unsrt}

\end{document}